\def\year{2021}\relax
\documentclass[letterpaper]{article} 
\usepackage{aaai21}  
\usepackage{colortbl}
\usepackage{hhline}
\usepackage{times}  
\usepackage{helvet} 
\usepackage{courier}  
\usepackage{float}
\usepackage[hyphens]{url}  
\usepackage{graphicx} 
\usepackage{colortbl}
\usepackage[table,xcdraw]{xcolor}
\usepackage{natbib}  
\usepackage{caption} 
\title{From the Head or the Heart? An Experimental Design on the Impact of Explanation on Cognitive and Affective Trust }

\author {
    Qiaoning Zhang,
    X. Jessie Yang, 
    Lionel P. Robert Jr. \\
}
\affiliations {
    University of Michigan \\
    qiaoning@umich.edu, xijyang@umich.edu, lprobert@umich.edu
}

\begin{document}

\maketitle

\begin{abstract}
Automated vehicles (AVs) are social robots that can potentially benefit our society. According to the existing literature, AV explanations can promote passengers’ trust by reducing the uncertainty associated with the AV’s reasoning and actions. However, the literature on AV explanations and trust has failed to consider how the type of trust—cognitive versus  affective—might alter this relationship. Yet, the existing literature has shown that the implications associated with trust  vary widely depending on whether it is cognitive or affective. To address this shortcoming and better understand the impacts of  explanations on trust in AVs, we designed a study to investigate the effectiveness of explanations on both cognitive and affective trust. We expect these results to be of great significance in designing AV explanations to promote AV trust.

\end{abstract}

\section{Introduction}
Automated vehicles (AVs) hold the potential for safer high- ways and reduced pollution; yet a lack of trust could hinder their adoption \cite{azevedo2020context}. AVs defined as the SAE levels 3–5 automated driving systems are capable of navigating roadways and interpreting traffic-control devices with little or no human involvement  \citep{taeihagh2019governing,gehrig1999dead}. Delegating most or all vehicle driving responsibilities to AVs can increase road safety and reduce the pollution emitted by vehicles \citep{duarte2018impact}. However, the general public has expressed growing skepticism about their safety and a lack of trust in the technology \citep{du2019look,robert2019automated,zhang_individual_2021}. Therefore, learning to promote trust in AVs remains a vital challenge.

Explanations—reasons or justifications for particular outcomes—have been shown to promote trust in AVs  \citep{du2019look,forster2017driver,ruijten2018enhancing,zhang2021drivers,haspiel2018explanations}. Explanations assist drivers with forming and strengthening a correct mental model, which makes the AVs’ actions predictable and understandable  \citep{korber2018have}. In addition, the automation transparency promoted by AV explanations can help drivers create an approximate representation of the system’s functions and competence in their mind, take the appropriate precautions in sudden takeover scenarios, understand the AV functions’ future actions, and trust the AV appropriately  \citep{forster2017increasing,toffetti2009citymobil,du2019look,du2020predicting}.

Despite the progress toward understanding the impact of explanations on AV trust, the literature has not considered how the type of trust—cognitive (rational-oriented) versus affective (emotional-oriented)—might alter this relationship. Literature examining interpersonal relationships has highlighted the importance of recognizing and incorporating  the distinction between cognitive and affective trust \citep{lewis1985trust,mcallister1995affect,robert2016monitoring}. Given the importance of different types of trust on interpersonal relationships, we seek to explore whether AV explanations affect cognitive and affective trust differently and how that might in turn influence human–AV interaction and AV adoption.

To better understand how the type of trust might influence the impact of the AV explanation on trust, we propose an experimental study employing a within-subjects design. Theoretically, this proposed research could provide insights and con tribute to the literature on AV explanations and trust. In addition, the results of this study have the potential to help the  design of AVs (1) consistently and effectively promote AV trust and (2) avoid situations where such designs are likely to fail.

\section{Background}
\subsection{Trust in Automated Vehicles and Explanation}

Trust has become a growing topic of interest in the domain of AVs  \cite{azevedo2021unified,azevedo2020real,petersen2019situational}. Defined as ``the willingness of a party to be vulnerable to the actions of another party based on the expectation that the other will perform a particular action important to the trustor, irrespective of the ability to monitor or control that other party." \citep[p.~712]{mayer1995integrative}, ), trust is a major construct for generating appropriate attitudes and predicting AV adoption.

“Explanation” is the reason the AV provides to the pas- senger to make its actions clear or easy to understand, and its impact on AV trust has been investigated in prior literature \cite{zhang2021drivers}. For example, Forster, Naujoks, and Neukum (2017) investigated the impact of AV explanation on trust based on Lee and See’s (2004) trust theory \citep{lee2004trust}. Forster et al. (2017) found that adding an explanation (i.e., semantic speech output) was superior to the no-explanation condition in terms of all trust dimensions. Also, Du et al. (2019) investigated the effect of explanation timing on trust that consisted of six dimensions based on the theoretical notion of Rempel, Holmes, and Zanna (1985) and Barber (1983). These dimensions included competence, predictability, dependability, responsibility, reliability, and faith \citep{barber1983logic,rempel1985trust}. Evidence  indicates that providing explanations before the AV takes ac tion can significantly increase trust compared to the conditions where either no explanation is provided or one is provided after the AV takes action. Ruijten, Terken, and Chandramouli (2018) and Hatfield (2018) investigated trust on the basis of Sheridan’s (1989) trust theory (i.e., familiarity, reliability, confidence) \citep{sheridan1989trustworthiness}. Similar to the result of Forster, Naujoks, and Neukum (2017), Ruijten et al. found that providing an explanation can effectively promote trust in AVs compared to no explanation. Results from Hatfield (2018) showed that the transparency produced by providing an explanation did not influence trust during forced moral outcomes (i.e., utilitarian or non-utilitarian decisions) \citep{hatfield2018effects}. 

In sum, trust is a major factor in the AV-related research. The impact of explanation on AV trust has been examined on the basis of several trust theories. Future studies are needed to more deeply explore AV trust to enhance and promote drivers’ trust in AVs.


\subsection{Cognitive and Affective Trust}

Previous literature on AVs has overlooked the distinction between two types of trust: affective and cognitive. Literature examining interpersonal trust has highlighted the importance of recognizing and incorporating the distinction between affective and cognitive trust  \citep{lewis1985trust,mcallister1995affect}. 

Cognitive trust is based on a cognitive or rational process that discriminates among trustworthy agents, distrusted agents, and unknown agents  \citep{lewis1985trust}. In interpersonal relationships, cognitive trust is when “we choose whom we will trust in which respects and under what circumstances, and we base the choice on what we take to be ‘good reasons,’ constituting evidence of trustworthiness”  \citep[p.~970]{lewis1985trust}. On this ground, people are associated with an experiential and rational process and ‘‘trust” cognitively by identifying reasons to trust and constituting evidence of trustworthiness.

Affective trust is complementary to cognitive trust and consists of an emotional bond among all those who participate in the relationship (i.e., feelings and emotions toward an object/agent) \citep{lewis1985trust}. In affective trust, people make emotional investments in trust relationships, express genuine care and concern for the welfare of partners, believe in the intrinsic virtue of such relationships, and believe that these sentiments are reciprocated \citep{pennings1987typology,rempel1985trust,mcallister1995affect}. 

Although little research, if any, has measured trust in AVs in light of the Lewis and Weigert (1985) trust theory, researchers in the domain of human–computer interaction have endeavored to describe and explain human–computer trust (HCT) from the perspectives of cognition and affect. Madsen and Gregor (2000) found that the overall perceived trust that a user has in a computer system (i.e., intelligent decision aid) comprises both cognitive and affective trust and that the affect-based trust is the strongest indicator of overall perceived trust in computer systems. \citep{madsen2000measuring}.

The objective of our proposed study is to investigate the effect of AV explanation timing on both cognitive and affective trusts. AV explanation timing is one important independent variable (i.e., explanation before/after the AV acts), and its impact on trust has been investigated in detail. However, its influence on trust has not been explored from the perspective of cognitive trust and affective trust. In this study, we seek to answer the following research question: Does the timing of the AV explanation influence the two types of trust—cognitive-based trust and affect-based trust?

\section{Methodology}

This study will employ a within-subjects experimental design on an online survey platform. The following subsections provide details about the proposed study.

\subsection{Participants}
The population to be examined will be U. S. drivers. All participants will be screened using a questionnaire for inclusion criteria. Participants must have a valid driver’s license. Before beginning the study, we will ask the university’s institutional review board to review and approve this study design.

We will perform a statistical power analysis to estimate sample size. The effect size (ES) in this study will be set based on data from a pilot study using Cohen, Cheung, and Raijman’s (1988) criteria \citep{cohen1988reliability}. With alpha = .05 and power = 0.80, the projected sample size needed with this effect size (GPower 3.1) can be derived for this within-group comparison.

\subsection{Study Design}
A within-subjects study with three conditions (i.e., no explanation, explanation before the AV acts, or explanation after the AV acts) will be employed to examine the research question. The sequence of these three AV explanation conditions will be counterbalanced via a Latin square design among participants. Each AV explanation condition will involve three unexpected and unique events differentiated by the driving environments and the actions of other vehicles on the roadway (i.e., events by other drivers; events by police vehicles; and events of unexpected re-routes in urban, highway, and rural environments). These three unexpected events will be based on previous literature  and correspond to realistic situations in automated driving \citep{du2019look,koo2015did,koo2016understanding}. Table 1 shows examples of the explanations to be provided by the AV.

\begin{table}
\centering
\caption{Event and Explanation Examples}
\begin{tabular}{cl} 
\hline
\rowcolor[rgb]{0.753,0.753,0.753} \textbf{Event} & \multicolumn{1}{c}{\textbf{Explanation}}                                                      \\ 
\hline
Heavy Traffic Rerouting                          & \begin{tabular}[c]{@{}l@{}}“Rerouting, traffic reported \\ahead."\end{tabular}                \\
Oversized Vehicle Ahead                          & \begin{tabular}[c]{@{}l@{}}“Oversized vehicle blocking \\roadway. Slowing down.”\end{tabular}  \\
Police Vehicle Approaching                       & \begin{tabular}[c]{@{}l@{}}“Emergency vehicle \\approaching. Stopping.”\end{tabular}          \\
\hline
\end{tabular}
\end{table}

\subsubsection{Independent Variables}

This study will use a within- subjects experimental design with the AV explanation timing as the independent variable. The AV will provide no explanation about its actions to the driver under the “no explanation” condition. The condition of “AV explanation before action” will involve the AV providing explanations prior to taking actions. For the “AV explanation after action” condition, the AV will provide explanations after the AV has taken actions.

\subsubsection{Dependent Variables}

This study’s dependent variables will be the cognitive- and affect-based trusts in AVs as measured by a 7-point Likert scale ranging from 1 (strongly disagree) to 7 (strongly agree) and adapted and from McAllister (1995). The items will be modified to suit the AV context. An example item is: “If people knew more about the automated vehicle, they would be more concerned and monitor its performance more closely.” To validate the questions, we will first conduct a pilot study and then run the principal component analysis (PCA) to identify underlying components and check the internal consistency of the data.

\section{Discussion}
We expect the results of this study to contribute to previous literature in the following ways. First, the findings of this study are expected to help explain and highlight the importance of explanation in the context of AVs.

Second, this proposed research should contribute to the prior literature by highlighting the distinction between cognitive-based and affect-based trusts in AVs as related to the explanation provided by AVs. As current research investigates the trust in AVs based on different trust theories, little research, if any, has recognized or investigated the trust in AVs from the perspective of trust types. Previous literature discussed the close relevance between trust and AV adoption \citep{jayaraman2019pedestrian}. By exploring the different types of trust, this paper can also uncover the ways in which AV adoption might be impacted by cognitive- and affect-based trusts. Practically, understanding more about the different types of trust and their impacts would be helpful to guide AV design toward consistently and effectively moderating the human–AV interaction and avoiding situations where such collaborations are likely to fail.

Third, we expect that the results of this study will direct future research to explore the factors that lead to different types of trust. Cognitive trust has been labeled “trust from the head” and is rooted in one person’s rational and objective assessment. The factors impacting people’s rational processes and giving them a reason to believe in AVs might influence cognitive trust. On the other hand, affective trust, labeled as “trust from the heart,” has  a more relational orientation and is closely associated with emotional exchanges and reciprocated sentiments. If a factor inspires people’s emotional desire to invest in AVs, then this factor might impact people’s affective trust.


\end{document}